\documentclass[english]{svproc}
\usepackage[T1]{fontenc}

\usepackage{color}
\usepackage{textcomp}
\usepackage{graphicx}
\usepackage{hyperref}
\usepackage{multirow}
\usepackage{multicol}
\usepackage{footmisc}
\usepackage{booktabs}

\makeatletter
\usepackage{graphicx}
\usepackage{caption}
\usepackage{subcaption}
\usepackage{siunitx}
\usepackage{amsmath}
\graphicspath{figures}

\makeatother

\usepackage{wrapfig}
\usepackage{babel}
\usepackage[style=ieee]{biblatex} 
\usepackage{authblk}

\begin{document}


\title{Carbon Footprint Evaluation of Code Generation through LLM as a Service}

\author{
  Tina Vartziotis\textsuperscript{1,5},
  Maximilian Schmidt\textsuperscript{1},
  George Dasoulas\textsuperscript{2},
  Ippolyti Dellatolas\textsuperscript{3},
  Stefano Attademo\textsuperscript{1},
  Viet Dung Le\textsuperscript{4},
  Anke Wiechmann\textsuperscript{4},
  Tim Hoffmann\textsuperscript{4},
  Michael Keckeisen\textsuperscript{1},
  Sotirios Kotsopoulos\textsuperscript{5}\protect\\
\vspace{4mm}
  \textsuperscript{1}TWT GmbH Science \& Innovation, Stuttgart, Germany \protect\\
  \textsuperscript{2}Harvard University \protect\\
  \textsuperscript{3}MIT \protect\\
  \textsuperscript{4}Mercedes-Benz \protect\\
  \textsuperscript{5}National Technical University of Athens, Greece
}

\authorrunning{Vartziotis et al. 2024}

\maketitle

\begin{abstract}
Due to increased computing use, data centers consume and emit a lot of energy and carbon. These contributions are expected to rise as big data analytics, digitization, and large AI models grow and become major components of daily working routines. To reduce the environmental impact of software development, green (sustainable) coding and claims that AI models can improve energy efficiency have grown in popularity. 
Furthermore, in the automotive industry, where software increasingly governs vehicle performance, safety, and user experience, the principles of green coding and AI-driven efficiency could significantly contribute to reducing the sector's environmental footprint. We present an overview of green coding and metrics to measure AI model sustainability awareness. This study introduces LLM as a service and uses a generative commercial AI language model, GitHub Copilot, to auto-generate code. Using sustainability metrics to quantify these AI models' sustainability awareness, we define the code's embodied and operational carbon.

\end{abstract}

\section{Introduction}
\subsection{Impact of AI on the Environment}
With widespread interest in using artificial intelligence and machine learning (ML) for a variety of applications, the environmental impact of AI and the ability to supply the electricity necessary to support its expansion have become growing concerns. Data center electricity consumption (excluding cryptocurrency mining) currently only accounts for 1-1.3\% of global electricity and 0.3\% of global carbon emissions \cite{iea2023}, \cite{masanet2020recalibrating}, \cite{hintemann2020}. Yet data centers are already competing with other uses of electricity in certain countries. In 2022, data centers in Ireland consumed 20\% of the country's total electricity, triple the 2015 levels \cite{IrelandCSO2023}. This trend could be exacerbated by a rising demand for clean electricity to support the deployment of heat pumps, electric vehicles, and electrification \cite{iea2022} in a context where renewable energy deployment is not meeting data center energy demand \cite{acun2023}. While efficiency gains and hardware advances have thus far reduced the energy consumption of data centers, it is unlikely that these practices will manage to offset increases in AI's electricity requirements in the long-term.
Demand for AI and ML is indeed expected to rise, as evidenced by the release of OpenAI's ChatGPT, which reached 100 million users in only 2 days. ChatGPT is an example of a Large Language Model (LLM), a generative AI model that uses natural language processing to predict words or sentences based on a certain context. Within the lifecycle of these models is a first training phase, where the models learn how to make the predictions, followed by an inference phase, where the models are deployed and used for a variety of applications. While both of these phases generate carbon emissions, most literature has focused on the emissions of the training phase \cite{strubell-etal-2019-energy, Patterson2021, Schwartz2020}. 
Yet emissions from the inference phase are non-negligible, with Google reporting 60\% of its AI-related energy consumption being due to inference \cite{patterson_llm}. Estimates for ChatGPT's electricity use during inference yielded 564MWh/day \cite{semianalysis}, compared to 1287 MWh used to train GPT-3 \cite{Patterson2021}. A recent study found that if generative AI were to be incorporated in every Google search query, Google's electricity consumption associated with AI would be 29.3TWh \cite{devries2023}. As a comparison, Google's total electricity consumption was at 18.3TWh in 2021, of which 10-15\% was due to AI.


\subsection{LLMs for sustainable code generation}
Large Language Models can contribute to sustainable code generation, including through: \textit{1. Optimized Code Suggestions}~\cite{codex, wang-etal-2021-codet5}: LLMs can suggest more efficient algorithms or code structures that perform tasks with fewer computational resources. This leads to reduced energy consumption in running the software. \textit{2. Automated Code Refactoring}~\cite{shirafuji2023refactoring}: LLMs can refactor existing code to make it more efficient and sustainable. By optimizing code for performance, they can reduce the computational power required, thereby saving energy. \textit{3. Educating Developers}~\cite{educating_developers}: They can provide developers with best practices for writing energy-efficient code. This includes advice on efficient algorithms, data structures, and design patterns. \textit{4. Reducing Development Time}~\cite{codex, chen2023frugalgpt}: By assisting in faster code development and debugging, LLMs reduce the overall time and energy spent in the software development lifecycle.
\textit{5. Code Analysis for Sustainability}~\cite{correct_Code}: LLMs can analyze existing codebases to identify areas where energy efficiency can be improved. They can suggest changes that reduce the environmental impact of software. \textit{6. Enhancing Algorithm Efficiency}: They can aid in designing more efficient machine learning models and algorithms, which are significant in reducing the carbon footprint of AI systems. \textit{7. Resource Allocation Optimization}: They can help optimize cloud resource usage, ensure that computing resources are used efficiently, and reduce waste and energy consumption. \textit{9. Predictive Maintenance and Optimization}~\cite{ZONTA2020106889}: LLMs can predict maintenance needs and optimize operations in IoT and other data-driven environments, reducing resource usage and waste.

\subsection{Green Coding Practices}
The carbon footprint of software, measured in $\mathrm{tCO}_2e$, is the product of its energy use in kWh and the carbon intensity of the energy used. The term green coding has therefore been used to describe the fact of powering data centers with cleaner sources of energy \cite{Henderson2020}, reducing the energy demand of data center cooling \cite{Lazic2018}, or using carbon offsets to claim overall emissions reductions \cite{Patterson2021}. Further, optimizing cloud workloads, along with adjusting the timing and geographical location for running computational tasks, particularly AI-related, can lead to substantial energy savings, especially for extended tasks \cite{Dodge2022}. 
The utilization of energy-efficient hardware forms another cornerstone strategy \cite{Lacoste2019}. Additionally, the enhancement of algorithm efficiency, via modifications in code architecture, contributes significantly to this effort \cite{So2019}. Accurate reporting of AI carbon intensity helps in quantifying and managing the environmental impact \cite{Dodge2022}. A comprehensive approach that accounts for emissions throughout all stages of software development, from initial training to final deployment, is imperative for a thorough lifecycle analysis \cite{Dodge2022}. Lastly, the development of smaller, custom-tailored LLMs, or 'Tiny LLMs', provides a feasible alternative that retains necessary functionalities with a significantly reduced computational and environmental footprint. This study excludes variability in grid carbon intensity, PUE, or hardware efficiency. All evaluations are on the same hardware and grid, focusing on efficiency gains solely from the code itself, such as algorithmic structure and instruction efficiency.


\section{Carbon Footprint Taxonomy}

\begin{figure}[t]
    \centering
    \includegraphics[width=\textwidth]{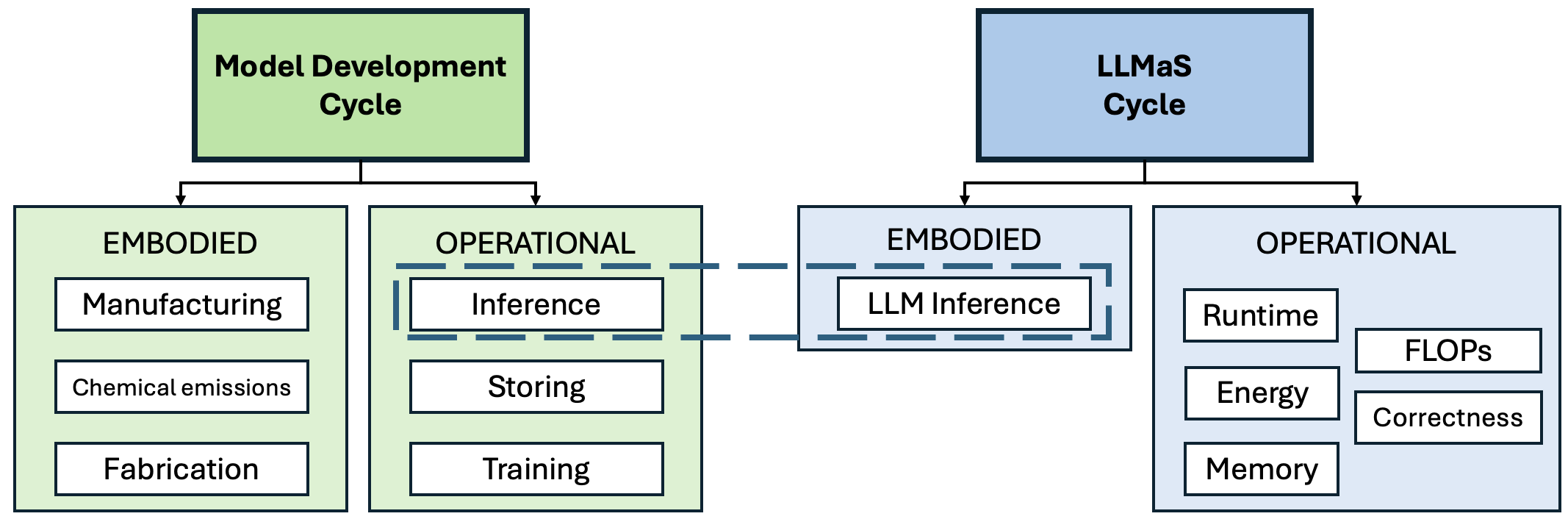}
    \caption{The cycle phases where operational and embodied carbon footprints occur. We study two production cycles: i) the Model Development Cycle (left), and ii) the LLMaaS Cycle (right). The embodied carbon footprint of the LLMaaS Cycle is overlapping in the LLM inference phase with the operational one by the Model Development Cycle.}
    \label{fig:cycles}
\end{figure}

We first make a distinction between two lifecycles of the LLM: i) the model development cycle, which corresponds to the infrastructure for LLM development, training, storing, and inference, and ii) the software engineering cycle, which makes use of the LLM as a service, a cycle that we call LLMaaS (LLM as a Service). We visualize the costs related to each cycle in Figure~\ref{fig:cycles}. The majority of the literature concentrates on the model development cycle~\cite{faiz2024llmcarbon, wu_2022_sustainable_ai}, assessing the embodied carbon footprint through hardware manufacturing costs, and the operational carbon footprint through the hardware utilization costs for training and inference.
\paragraph{Model Development Cycle.} The embodied carbon footprint includes all energy costs derived from hardware manufacturing. Specifically, previous studies report the Carbon emitted Per unit Area (CPA) for various computing and memory units, including GPUs, TPUs, CPUs, and SSDs~\cite{wu_2022_sustainable_ai, choe2021, faiz2024llmcarbon}. In all cases other than the CPA per computing chip, the total embodied carbon footprint depends on the execution duration and the lifespan of each chip.

The operational carbon footprint in the Model Development Cycle is computed from the energy costs of hardware utilization during the training and the inference phase, as well as the storing requirements of the model, and data parameters for the maintenance of the LLMs~\cite{patterson_llm, henderson_llm, dodge_2022}. In~\cite{faiz2024llmcarbon}, the authors introduce an Operational Carbon Model (OCM), that makes use of the FLOP count, the hardware efficiency, and the computing device number to estimate the execution time of a device, and consequently the consumed energy for all training, inference, and storing processes. Combining the introduced OCM, and the embodied carbon footprint estimation, the authors develop LLMCarbon, which estimates the total carbon footprint associated with the Model Development Cycle of LLMs.

\paragraph{LLM as a Service Cycle.} Moving beyond the Model Development Cycle, the software engineering pipeline that uses LLMs as a service does not consider the development and training of LLMs as part of the embodied computations and costs.
Specifically, the embodied carbon footprint of LLMaaS Cycle is directly related to the costs occurring from the LLM inference, as shown in Figure~\ref{fig:cycles}. The operational carbon footprint of the LLMaaS Cycle depends on the costs associated with the output source derived from the LLM. We have extensively discussed the latter costs in previous work~\cite{vartziotis2024learn}. In particular, we define the Green Capacity (GC) of the output source based on 5 sustainability metrics: a) the execution runtime, b) the required memory, c) the energy consumption, d) the total number of floating point operations (FLOPs), and e) the correctness of the output source. 

In the remaining part of this work, we concentrate on evaluating the embodied and operational carbon footprint of the LLMaaS Cycle, defined as follows:

\begin{enumerate}
    \item \textbf{Embodied Carbon:} Embodied carbon refers to the greenhouse gas emissions associated with the infrastructure required for the development of a code block. In the case of LLM-generated code, it corresponds to the LLM inference, while for human-generated code, it corresponds to the energy used by the hardware while developing the code. 
    
    \item \textbf{Operational Carbon:} Operational carbon corresponds to the emissions generated during the active utilization of a code block~\cite{vartziotis2024learn}. 

    \item \textbf{Dynamic Embodied Carbon:} Dynamic embodied carbon refers to the ongoing emissions that occur through the continuous development and updating of code. It accounts for the carbon footprint associated not just with the initial creation of software but also with its iterative improvements, modifications, and extensions over time. This concept recognizes the evolving nature of software and the need to consider sustainability in the persistent development cycle.
    \item \textbf{Carbon Intensity (CI)} Carbon intensity quantifies the grams of $CO_{2eq}$ emitted per kWh of energy used in a specific grid. The energy grid and its carbon intensity are defined for a given location based on the local energy sources~\cite{henderson2022systematic}. Specifically for Germany, the hourly carbon intensity is reported on the publicly available electricity maps\footnote{https://app.electricitymaps.com}. In the context of data centers, there are reported values for the carbon intensity of a specific data center, e.g., a Europe-based data-center has 91\% carbon-free energy and $CI = 127g CO_{2eq}/kWh $ compared to an East Asia-based one, that has 28\% carbon-free energy and $CI = 360g CO_{2eq}/kWh$~\cite{faiz2024llmcarbon}.
    \item \textbf{Carbon Footprint:} The carbon footprint is the amount of carbon emitted during a computational operation \cite{faiz2024llmcarbon}. Formally, it is defined as $CO_{2eq}= Energy * CI$. Estimating the carbon footprint is involves multiple parameters, which depend on the type of computational units used and their characteristics (e.g. local servers, large data centers, or local computers).
    \item \textbf{Embodied Energy (Inference):} The embodied energy during inference is the total energy consumption (in kWh) of the inference step. Measuring this value requires external power meters or server hardware support, and different organizations use various monitoring tools to evaluate it. As an example, Microsoft Fabric\footnote{https://learn.microsoft.com/en-us/fabric/enterprise/licenses}, an analytics solution designed by Microsoft, defines execution time statistics of the required computational units as Capacity Unit (CU) seconds.
\end{enumerate}

\begin{table}[t]
\centering
\caption{File Operations in Frontend and Backend Processes. For each process, the input corresponds to the source already written, which CoPilot has access to, while the output corresponds to the test modules that are generated. CR is an abbreviation for Consumption Rate, as this is measured by Microsoft Fabric.
}
\label{table:file-operations}
\resizebox{\textwidth}{!}{
\begin{tabular}{lccccc}
\hline
              & \textbf{Input Frontend} & \textbf{Output Frontend} & \textbf{Input Backend} & \textbf{Output Backend} \\ \hline
Files         & 192                     & 149                          & 208                        & 50                            \\   
Words/File    & 177                  & 80                           & 300                        & 150                           \\ 
Token/File    & 235             & 107                  & 400                        & 200                           \\
CR/File      & 94             & 128                          & 160                        & 240                           \\
CR           & 18073                 & 19072                        & 33280                      & 12000                         \\
Token         & 45184                   & 8533                         & 120000                     & 30000                         \\ \hline
\end{tabular}}
\end{table}
\section{Embodied Costs}

\subsection{Case Study in Software Testing}
This case study focuses on a cloud-based web application characterized by stringent web security requirements. We aim to demonstrate the significant improvements in operational efficiency and accuracy in code development achieved through integrating LLMs. The web application necessitates testing and comprehensive updates of the underlying software frameworks. This update requires code refactoring and, in some instances, complete rewriting due to non-backward compatible changes in Angular's testing framework. 

\textbf{Experimental Methodology} The application includes a broad range of functionalities, for example SQL database operations, UI/UX frontend visualizations, and collaborative requirements engineering. The backend is developed using Java with the Spring Boot framework, while the frontend employs TypeScript, HTML, and CSS with Angular. In this use case, we focused on generating 149 frontend and 50 backend software testing scenarios. The evaluation of the inference is estimated by the tokens used in this pipeline, both for the input prompt and the output completion. Copilot consumption is measured\footnote{https://learn.microsoft.com/de-de/fabric/get-started/copilot-fabric-consumption} by the number of tokens processed. Tokens can be thought of as pieces of words. Approximately 1,000 tokens are about 750 words. The energy cost varies based on the geographic region\footnote{https://learn.microsoft.com/en-us/fabric/data-science/ai-services/ai-services-overview}. 


Tables \ref{table:file-operations} and \ref{table:consumption-rates} provide a detailed evaluation of GitHub Copilot's impact in a software testing and refactoring context. These elements compare the estimated tokens for task completion with GitHub Copilot. Specifically, in Table~\ref{table:file-operations}, we show the statistics regarding the input prompt files and the output testing files both for the frontend and backend scenarios, including the ratios $\frac{\text{Words}}{\text{File}},\frac{\text{Tokens}}{\text{File}}$ and Consumption Rate (CR), measured in seconds. For the latter's values, we follow the reports from Microsoft Fabric solution, that introduce and monitor the Capacity Unit (CU) time measurement (measured in seconds).



\setlength{\intextsep}{0pt}
\begin{wraptable}{R}{45mm}
\centering
\caption{Consumption Rates (Seconds/Token)}
\label{table:consumption-rates}
\begin{tabular}{lc}
\hline
              & \textbf{CR(s)/Token}  \\ \hline
Input  & 0.4                  \\
Output  & 1.2                \\\hline
\end{tabular}
\end{wraptable}


\textbf{Time Savings} The total time consumption highlights the significant time savings offered by GitHub Copilot. While completing all tasks without the tool was estimated to require approximately 78 hours, using GitHub Copilot reduced this time to just over 5 hours. Despite some limitations in complex backend tasks, this represents a substantial increase in efficiency.

 The embodied carbon of Copilot can not be accurately estimated due to the lack of information. In order to provide an understanding of the magnitude of the carbon emissions, we will estimate the LLM inference based on the workload in a server as provided by Intel. As defined \[{CO_{2eq}}^{(Emb)}= Energy_{emb} * CI,\]  the carbon footprint is computed based on the energy to generate the prompting coding results multiplied by the carbon intensity (CI) of the data center. The magnitude is shown in Table \ref{table:emissions_energy}, which summarizes the energy consumed during the code's development and operation.
 More specifically, the energy based on Intel's simplified computational method\footnote{The method has been presented in a blog post by Intel on \hyperlink{https://medium.com/intel-analytics-software/reduce-large-language-model-carbon-footprint-with-intel-neural-compressor-and-intel-extension-for-dfadec3af76a}{Medium}.} to provide a rough estimate of the real emissions \[Energy_{emb} = P * T * N \]  where $P$ is the server power consumption (in Watts), $T =0,47s$ is token latency (in seconds), which refers to the time it takes for the LLM to process a single token during inference, and $N= 203717$ is the number of tokens (input and output) processed during inference. The CPU and memory subsystem are the two major components contributing to server power consumption: $P = P_{cpu} + P_{mem} * M= (350W+ 0.1W/GB* 60GB),$ where $P_{cpu}$ is the CPU power consumption, $P_{mem}$ is the power consumption per unit of memory (in Watts/GB), and $M$ the memory usage. 


\section{Operational costs}

In addition to the inherent costs of software development, the costs incurred during code execution also contribute to the overall environmental impact made by the pipeline. We call the latter the "greenness" of code. These operational costs arise from running the application's source code or any supporting code (e.g. software testing) throughout the software's lifecycle. In the following sections, we will first examine the metrics, methodology, and tools used to evaluate the operational costs of the previously discussed use case. We will then analyze the operational cost of the software testing case study. 

\subsection{Sustainability Metrics for Software Operation}
Before any statements regarding the greenness of code can be made, we must first define a set of metrics that capture the characteristics of the code that contribute to its greenness. In this paper, we use the metrics that we defined in our previous work~\cite{vartziotis2024learn}. The metrics and a brief description are listed below: \textit{1. Code Correctness}: defines whether the code yields valid results. \textit{2. Runtime}: measures the time required by the code to execute the required computations, e.g., start to end for algorithmic computations. \textit{3. Memory}: corresponds to the peak amount of memory reserved by the application during execution. \textit{4. FLOPs}: is the total amount of executed floating-point operations. \textit{5. Operational Energy}: energy consumed during the execution of the code. For further details, we refer to our previous publication~\cite{vartziotis2024learn}.

Although the metrics we identified can generally be used for every code or application, different types of applications or code may require changes in the way metrics are applied and results interpreted.
For example, a web application's backend usually runs 24 hours a day on a server. Thus, the runtime metric does not provide beneficial information about the code's sustainability. Similarly, the peak memory usage of a web application may vary a lot depending on the number of users of the application. If a large number of users access the application at the same time, more memory might be required for processing the requests or caching data than in cases where only a few users access the application. Regarding FLOPs, if an application does not perform any floating-point operations, this metric can be left out completely.

\subsection{ Operational Costs of Case Study in Software Testing}

In the next case study, we evaluate the energy consumption and runtime of the test suite written using GitHub Copilot. For the operational carbon of tests for the source code, we limit the metrics we record to the energy consumption during the execution of the tests.  We also evaluated the code correctness to validate the correctness of the actual code, meaning that the tests must be correct; otherwise, the whole execution would be flawed. The tests for the frontend application were all successfully generated based also on their low complexity while the tests for the backend applications had a 50\% accuracy rate, which required a software engineer to correct and rewrite part of the generated code. The other metrics do not impact the carbon footprint and are thus not considered in this scenario.

The tests' memory usage in this scenario is very small since the memory usage during test execution does not represent the real environment of the application. The test data focuses on special cases inside the data to test the functionality. The evaluation of the FLOPs was excluded in this case study since the application performs a negligible amount of floating-point operations.

There are numerous tools available to evaluate energy consumption. The most common and efficient Linux-based tool is \texttt{perf}~\cite{vartziotis2024learn}. Based on the platform and technology stack available for this case, we chose a Windows operating tool, \textit{Windows Energy Estimation Engine} (short E3), to evaluate the energy consumption of the testing codes. E3 is a monitoring tool built into the Windows OS to monitor the energy consumption of all running processes and the different hardware, such as the CPU or hard drives.
One major limitation of E3 is that it is only available on Windows devices that run on a battery. 

E3 records all running processes which is why we must execute the testing code in a separate terminal on our computer to be able to identify it in the list of records E3 creates.
Furthermore, E3 stores records of the processes' energy consumption in one minute intervals meaning that the resulting CSV file contains multiple entries for a process if it is running longer than one minute.
Therefore, we filter the list of recordings generated by E3 to only include our specific process and then sum up the values E3 stores in the \textit{TotalEnergyConsumption} column.
The sum of these values for our specific process represents the total energy consumption of our code during the execution including the energy consumption of the CPU, disks (if used) and the other hardware involved.

\textbf{Operational Carbon Footprint: } Using the same approach for computing the carbon emission, we result in eight times less than the equivalent embodied carbon emissions \[{CO_{2eq}^{(Oper)}}= Energy_{op} * CI,\] resulting in a \textbf{total Carbon LLMaaS Footprint:  \[{CO_{2eq}^{(LLMaaS)}}={CO_{2eq}^{(Oper)}} + {CO_{2eq}^{(Emb)}} \]}.


\begin{table}[h!]
\centering
\caption{Operational Carbon Emissions}
\begin{tabular}{lcc}
\hline
Oper. Energy & kJ & kWh \\
\hline
Frontend & 0.446 & 0.0001 \\
Backend & 4190.5 & 1.1314 \\
\hline
\hline
\end{tabular}
\label{table:carbon_emissions}
\end{table}


\begin{table}[h!]
\centering
\caption{Summary of Carbon Emissions and Energy Consumption}
\begin{tabular}{l r}
\hline
\textbf{Metric} & \textbf{Value} \\
\hline
Embodied Energy & 9.203 kWh \\
Operational Energy & 1.131 kWh \\
Carbon Intensity & 0.172 kgCO2e/kWh \\
Embodied Carbon Emissions & 1.582 kgCO2e \\
Operational Carbon Emissions & 0.194 kgCO2e \\
Total Carbon Emissions $(LLMaaS_{CO_{2eq}})$ & 1.777 kgCO2e \\
\hline
\end{tabular}
\label{table:emissions_energy}
\end{table}


\section{Conclusion}

Our research investigates the environmental implications of AI-driven code generation, specifically looking at Large Language Models (LLMs) offered as a service for software development. We examine both the embodied carbon footprint from running inference on massive LLMs, as well as the operational emissions from executing the output code generated by the trained LLMs.

We find that the capability of AI assistant tools (e.g., GitHub Copilot) to improve energy efficiency is strongly dependent on the various phases of the embodied and operational costs. This highlights the growing need to adopt green coding practices and efficient management of computing resources throughout the LLM as a Service (LLMaaS) Cycle. Measuring and mitigating the ecological footprint of LLMaaS proves complex. Robust sustainability metrics and continued greening of AI technologies themselves are needed. It is essential to thoroughly examine the full environmental impact of utilizing Large Language Models as a Service (LLMaas) and its influence on the broader coding community. Currently, the methods and tools available to assess the environmental footprint of computing activities are inadequate for fostering the necessary awareness. This is especially relevant for industries like automotive, where software plays an increasingly large role in vehicle performance and sustainability.

\section{Acknowledgments}
This work is inspired by a development project between Mercedes-Benz and TWT. We express our gratitude to IBM, Microsoft, and GitHub for the discussions on automated code generation that we had as part of this project.
We are grateful to Florian Schneider, Tobias Roessler, and Laura Casella from TWT, our collaborators NIKI Ltd Digital Engineering, and Mercedes-Benz for their engaging and informative discussions, which significantly influenced the project’s scope.

\printbibliography
\end{document}